\documentclass[sigconf, table]{acmart}

\copyrightyear{2018} 
\acmYear{2018} 
\setcopyright{acmcopyright}
\acmConference[CCS '18]{2018 ACM SIGSAC Conference on Computer and Communications Security}{October 15--19, 2018}{Toronto, ON, Canada}
\acmBooktitle{2018 ACM SIGSAC Conference on Computer and Communications Security (CCS '18), October 15--19, 2018, Toronto, ON, Canada}
\acmPrice{15.00}
\acmDOI{10.1145/3243734.3243808}
\acmISBN{978-1-4503-5693-0/18/10}

\usepackage{epsfig,endnotes}

\usepackage{natbib}
\setcitestyle{numbers,sort&compress}

\usepackage[normalem]{ulem}
\usepackage{longtable}
\usepackage{xcolor}
\usepackage{tabularx}
\usepackage{pdflscape}
\usepackage{amsmath}

\usepackage{adjustbox}
\usepackage{changepage}
\usepackage{hhline}
\usepackage{pbox}
\usepackage{graphicx}
\usepackage{mdframed,enumitem}

\usepackage{algorithmic}
\usepackage[algoruled]{algorithm2e}

\usepackage{subfigure}
\usepackage{hyperref}
\usepackage{multirow}
\usepackage{threeparttable}

\usepackage[T1]{fontenc}
\usepackage{listings}
\usepackage{color}
\usepackage{balance}

\newcommand{\myparatight}[1]{\smallskip\noindent{\bf {#1}:}~}
\newenvironment{packeditemize}{\begin{list}{$\bullet$}{\setlength{\itemsep}{0.2pt}\addtolength{\labelwidth}{-4pt}\setlength{\leftmargin}{\labelwidth}\setlength{\listparindent}{\parindent}\setlength{\parsep}{1pt}\setlength{\topsep}{0pt}}}{\end{list}}

\definecolor{mygreen}{rgb}{0,0.6,0}
\definecolor{mygray}{rgb}{0.5,0.5,0.5}
\definecolor{mymauve}{rgb}{0.58,0,0.82}
\usepackage{amsmath}
\usepackage{amssymb}
\usepackage{pdfrender}
\usepackage{pifont}

\begin{document}
\title[EviHunter]{EviHunter: Identifying Digital Evidence in the Permanent Storage of Android Devices via Static Analysis}

\author{Chris Chao-Chun Cheng, Chen Shi, Neil Zhenqiang Gong, and Yong Guan}
\affiliation{Department of Electrical and Computer Engineering\\ NIST Center of Excellence in Forensic Science - CSAFE\\ Iowa State University}
\affiliation{\{cccheng, cshi, neilgong, guan\}@iastate.edu}

\begin{abstract}
Crimes, both physical and cyber, increasingly involve smartphones due to their ubiquity. Therefore, digital evidence on smartphones plays an increasingly important role in crime investigations. Digital evidence could reside in the memory and permanent storage of a smartphone. While we have witnessed significant progresses on memory forensics recently, identifying evidence in the permanent storage  is still an underdeveloped research area. Most existing studies on permanent-storage forensics rely on manual analysis or keyword-based scanning of the permanent storage. Manual analysis is costly, while keyword matching often misses the evidentiary data that do not have interesting keywords.   

In this work, we develop a tool called \emph{EviHunter} to automatically identify evidentiary data in the permanent storage of an Android device.  There could be thousands of files on the permanent storage of a smartphone. A basic question a forensic investigator often faces is which files could store evidentiary data. EviHunter aims to answer this question. Our intuition is that the evidentiary data were produced by apps; and an app's code has rich information about the types of data the app may write to a permanent storage and the files the data are written to. Therefore, EviHunter first pre-computes an \emph{App Evidence Database (AED)} via static analysis of a large number of apps. The AED includes the types of evidentiary data and files that store them for each app. Then, EviHunter matches the files on a smartphone's permanent storage against the AED to identify the files that could store evidentiary data. 
 We evaluate EviHunter on benchmark apps and 8,690 real-world apps.  Our results show that EviHunter can precisely identify both the types of evidentiary data and the files that store them. 

\end{abstract}



\maketitle

\section{Introduction}
\label{sec:introduction}



Smartphones are playing an increasingly important role in investigating both cyber and physical crimes, as they are pervasive devices and they capture both online and offline activities of their owners. 
For instance, even several years ago when smartphones were not as pervasive as today, the number of crimes that involve mobile-phone evidence increased 10\% per year on average from 2006 to 2011~\cite{crimeMobileEvidence}. 
In 2017, a visiting scholar at University of Illinois Urbana-Champaign was kidnapped. Via forensic examination of the suspect's smartphone, FBI agents found that the smartphone owner visited websites about ``perfect abduction fantasy'' and ``planning a kidnapping'', which has been used as important digital evidence investigation~\cite{uiucCaseComplaint}. Likewise, in one of the biggest poaching cases in Washington~\cite{gpsCasePoach}, suspects took pictures and videos of killing animals (illegally). GPS locations retrieved from these images and videos led to dozens of kill sites where physical evidence (e.g., bullets) was found, and texts retrieved from the suspects' smartphones were used as digital evidence to corroborate charges.

Digital evidence on a smartphone could reside in the memory or permanent storage, e.g., flash storage, SD card. Recently, a series of tools~\cite{DSCRETE,VCR,GUITAR,RetroScope} have been developed to significantly advance memory forensics. However, identifying digital evidence in the permanent storage of a smartphone is still an underdeveloped research area. While memory forensics can reconstruct a smartphone's context that is probably in a short period of time before the last use of the smartphone,  permanent-storage forensics could uncover evidentiary data about the smartphone's historical activities over a long period of time, as some criminal cases may be planned and conducted across a large range of dates and locations. In this work, we focus on permanent-storage forensics. More specifically, data are stored in files by apps on a permanent storage; and one basic problem a forensic investigator often faces is which files, among the possibly thousands of files on a suspect's smartphone, could store relevant evidentiary data (e.g., GPS locations, visited URLs, and texts). We call this problem \emph{evidence identification problem}. After identifying such files, forensic investigators can then retrieve, decode, or reconstruct the evidentiary data from them. 

Existing studies on this evidence identification problem mainly rely on manual analysis~\cite{al2013skype,alyahya2017snapchat,anglano2014forensic,wu2017forensic} or keyword-based scanning~\cite{daryabar2016automated,UFED,XRY,FTK}. For instance, in keyword-based scanning, a file whose file name or content includes the keywords GPS, latitude, or longitude is labelled as a file that could include GPS locations. Manual analysis is time-consuming and error-prone, while keyword-based scanning misses the files that do not include the specified keywords. Indeed, studies~\cite{mfHans2017} showed that keyword-based scanning can only identify a small fraction of files that could store evidentiary data.

\myparatight{Our work} In this work, we develop \emph{EviHunter}, a tool to automatically identify the files on an Android device that could store the evidentiary data of interest to forensic investigators. Our intuition is that the files/data on a smartphone were produced by apps; an app's code contains rich information about 1) the types of data that the app could write into a file system and 2) the files where the data are written to. 

Based on the intuition, EviHunter takes an \emph{offline-online} approach. In the offline phase, EviHunter builds an \emph{App Evidence Database (AED)} for a large number of apps via static data flow analysis. Specifically, for each app, the AED includes the files that could store evidentiary data of interest and the types (e.g., location, visited URL, text input, and time) of evidentiary data in each of such files, where a file is represented using its complete \emph{file path} on Android. We adopt static analysis instead of dynamic analysis to have high coverage and be less likely to miss the files that could contain evidentiary data. In the online phase, given a smartphone, EviHunter matches the file paths on the smartphone's file system against those in the AED to identify the files that could contain evidentiary data. 

Building the AED is a key challenge for EviHunter. We note that a large number of static analysis tools--such as CHEX~\cite{CHEX}, FlowDroid~\cite{flowdroid}, AmanDroid~\cite{Amandroid}, DroidSafe~\cite{droidsafe}, R-Droid~\cite{rDroid}, IccTA~\cite{iccta}, and HornDroid~\cite{HornDroid}--have been developed to detect sensitive data flows between sources and sinks in Android apps. These tools were designed to detect the data that could flow from certain \emph{sources} to \emph{sinks}, where a source is where data are created and a sink is where the data ends, 
e.g., file system is a sink in our problem. However, these tools did not consider the files where the data are written to. 
For instance, these tools could detect that an app will collect GPS locations and save them to the file system, but they do not report the files where the GPS locations will be written to. One possible reason is that these tools were designed for security and privacy purposes; and it does not matter much which files sensitive data are written to in terms of privacy leakage.

To address the challenge, EviHunter extends existing static data flow analysis techniques for Android in several aspects. First, in some existing static analysis tools, a tag is associated with a variable to represent the types of sensitive data in the variable. We extend the tag to include both the types of sensitive/evidentiary data and the file path associated with a variable. Second, we extend the propagation rules to spread both the types of evidentiary data and file paths as we analyze the statements in an app. Third, we leverage techniques developed by HornDroid~\cite{HornDroid} to partially address multi-threading and reflection. 
Fourth, for complexity consideration, we manually summarize the semantics for the commonly used system APIs as some previous tools did~\cite{CHEX,flowdroid,Amandroid}. However, we further summarize the semantics of the system APIs for file path construction, which were not considered by previous tools. Fifth, we extend the sources by 1) uncovering the source methods for file path construction (these were not considered by previous tools), and 2) uncovering new sources for sensitive/evidentiary data including location, visited URL, and time (these sources were considered by previous tools but not complete).

In implementing EviHunter, we leverage Soot to transform an Android app to Jimple code (a three address intermediate representation), IC3~\cite{ic3} to build inter-connected component communication models, and FlowDroid~\cite{flowdroid} to build call graphs and entry points. Then, EviHunter performs forward analysis of the Jimple code. We evaluate EviHunter with respect to AED construction using 1) the benchmark apps in DroidBench~\cite{droidBench} that have file-system sinks, 2) new benchmark apps that we design to test the scenarios that are not covered by the DroidBench apps, and 3) 8,690 real-world apps. Our results show that EviHunter can precisely and accurately find the files that could store evidentiary data for benchmark apps. Moreover, we performed a best-efforts manual verification of the results for 60 randomly sampled real-world apps. Our results show that EviHunter achieves a precision of 90\% and a recall of 89\% at identifying the files that could contain evidentiary data. Although our work does not focus on privacy leakage, our results do have interesting privacy implications. In particular, we found that some apps save GPS locations, visited URLs, and text inputs on the external storage of an Android device. An app with the READ\_EXTERNAL\_STORAGE permission can access these data to track and profile the user. Finally, we use a case study to show how a forensic investigator can use EviHunter to find evidentiary data on a smartphone. 

In summary, our contributions are as follows:

\begin{packeditemize}
\item We develop EviHunter to automatically identify the files on an Android device that could contain evidentiary data. 
\item We extend existing static analysis tools for Android to detect both types of evidentiary data an app could write to file system and the files where the data are written to.
\item We evaluate EviHunter using both benchmark apps and real-world apps. Our results show that EviHunter achieves high precisions and recalls.  

\end{packeditemize}

\begin{table*}[!htbp]

\caption{Android APIs to get file paths.}
\label{table:fileAPI}
\vspace{2mm}

\centering
\bgroup
\def\arraystretch{1.2}
\makebox[\linewidth]{\small
\begin{threeparttable}[]

\addtolength{\tabcolsep}{-4pt}
\begin{tabular}{|c|l|}
\hline
\multicolumn{1}{|c|}{\textbf{API}} & \multicolumn{1}{c|}{\textbf{Description}} \\ \hline


\textit{getDataDirectory()} & Return \texttt{File} object whose path is  "/data/" \\ \hline

\textit{getCacheDir()} & Return \texttt{File} object whose path is "/data/data/<package name>/cache/"  \\ \hline

\textit{getFilesDir()} & Return \texttt{File} object whose path is "/data/data/\textless package name\textgreater/files/" \\ \hline

\textit{getFileStreamPath(String arg0)} & Return \texttt{File} object whose path is "/data/data/\textless package name\textgreater/files/\textless arg0\textgreater"\\ \hline

\textit{openFileOutput(String arg0, int arg1)} & Return \texttt{FileOutputStream} object whose path is "/data/data/\textless package name\textgreater/files/\textless arg0\textgreater" \\ \hline

\textit{getDataDir()} & Return \texttt{File} object whose path is "/data/data/\textless package name\textgreater/databases/" \\ \hline

\textit{openOrCreateDatabase(String arg0, ... )} & Return \texttt{SQLiteDatabase} object whose path is "/data/data/\textless package name\textgreater/databases/\textless arg0\textgreater" \\ \hline

\textit{openDatabase(String arg0, ...)} & Return \texttt{SQLiteDatabase} object whose path is "/data/data/\textless package name\textgreater/databases/\textless arg0\textgreater" \\ \hline

\textit{getSharedPreferences(String arg0,int arg1)} & Return \texttt{SharedPreferences} object whose path is "/data/data/\textless package name\textgreater/shared\_prefs/\textless arg0\textgreater.xml" \\ \hline

\multirow{2}{*}{\textit{getDefaultSharedPreferences(Context arg0)}} & Return \texttt{SharedPreferences} object whose path is \\ 
& \hspace{3cm}"/data/data/\textless package name\textgreater/shared\_prefs/\textless package name\textgreater\_preferences.xml" \\ \hline

\textit{getPreferences(int arg0)} & Return \texttt{SharedPreferences} object whose path is "/data/data/\textless package name\textgreater/shared\_prefs/\textless context\textgreater\tnote{1}  .xml" \\ \hline

\textit{getDir(String arg0, int arg1)} & Return \texttt{File} object whose path is "/data/data/\textless package name\textgreater/app\_\textless arg0\textgreater/" \\ \hline

\textit{openOrCreateDatabase(File arg0, ... )} & Return \texttt{SQLiteDatabase} object whose path is "\textless arg0\textgreater" \\ \hline


\textit{getExternalStorageDirectory()} & Return \texttt{File} object whose path is "/sdcard/" \\ \hline

\textit{getExternalStoragePublicDirectory(String arg0)} & 
\begin{tabular}[c]{@{}l@{}}
	Return \texttt{File} object whose path is "/sdcard/", if \textless arg0\textgreater is empty.\\ 
	Return \texttt{File} object whose path is "/sdcard/\textless arg0\textgreater/", otherwise.
\end{tabular} \\ \hline

\textit{getObbDir()} & Return \texttt{File} object whose path is "/sdcard/Android/obb/\textless package name\textgreater/" \\ \hline

\textit{getExternalCacheDir()} & Return \texttt{File} object whose path is "/sdcard/Android/data/\textless package name\textgreater/cache/" \\ \hline

\textit{getExternalFilesDir(String arg0)} & 
\begin{tabular}[c]{@{}l@{}}
Return \texttt{File} object whose path is "/sdcard/Android/data/\textless package name\textgreater/files/", if \textless arg0\textgreater is empty.\\ 
Return \texttt{File} object whose path is "/sdcard/Android/data/\textless package name\textgreater/files/\textless arg0\textgreater/", otherwise.
\end{tabular} \\ \hline


\end{tabular}
\vspace*{0.2cm}

\begin{tablenotes}
\item [1] \textless context\textgreater \ refers to an app's environment information, such as launching activity class name and service name. 
\end{tablenotes}

\end{threeparttable}}
\egroup

\end{table*}

\section{Background and Problem Formulation}
\label{sec:background}

We first briefly introduce the Android file system used to manage the permanent storage and then define our forensics problem. 

\subsection{Android File System}
\label{sec:filesystem}

\begin{figure}[]
\centering
\includegraphics[]{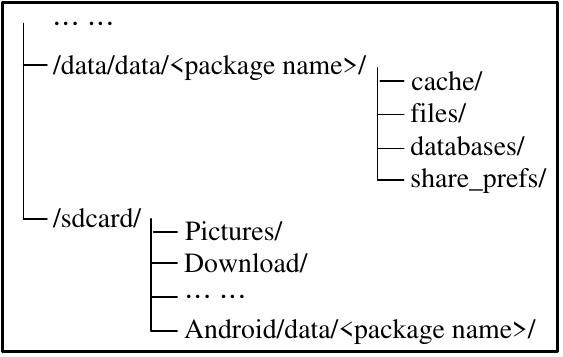}
\caption{Illustration of the \textit{/data/} and \textit{/sdcard/} directories in Android's file system.}
\label{structure}
\end{figure}

\myparatight{Directory structure} Android file system pre-defines several top-level directories, e.g., \textit{/data/} and \textit{/sdcard/}. The directory \textit{/data/} is on the internal storage of Android, while \textit{/sdcard/} is linked to external storage. Figure~\ref{structure} shows the structure of the \textit{/data/} and \textit{/sdcard/} directories.
Each app has a default directory \textit{/data/data/<package name>/}, where the package name is treated as the identifier of the app. For instance, the Facebook app's package name is com.facebook.katana and an Android device will create a directory \textit{/data/data/com.facebook. katana/} after the Facebook app is installed on the device. 
An app's directory stores various data of the app. Android also pre-defines several sub-directories under an app's directory. Example sub-directories include \textit{/files/}, \textit{/databases/}, \textit{/share\_prefs/}, and \textit{/cache/}. The sub-directory \textit{/files/} can include any files, \textit{/databases/} stores SQLite database files,  \textit{/share\_prefs/} stores SharedPreferences files, and \textit{/cache/} stores cache files. A SharedPreferences file can be viewed as a simple database, in which data are stored in key-value pairs. 

\begin{lstlisting}[float, frame=single,numbersep=-10pt,numbers=left,language=java,caption={An example of soft-coded file access.},label=list:motivateExample]
    foo(String gpsLong, String gpsLat){
        String fileName = "locSink";
        FileOutputStream sink 
        	= openFileOutput(fileName,0);
        long time = System.currentTimeMillis();
        String str = gpsLong + gpsLat + time;
        sink.write(str.getBytes());
        sink.close();
    }
\end{lstlisting}

When an app wants to use external storage and has the permission to do so, Android will create a directory \textit{/sdcard/Android/data/<package name>/} for the app. Moreover, this directory contains sub-directories \textit{/files/} and \textit{/cache/}.
Unlike the internal storage, data stored in the external storage can be read by all other apps on the device. Moreover,
the external storage also has other public directories such as \textit{Pictures} and \textit{Download}. We note that if a device does not have a real SD card, Android will use a part of the internal storage to emulate one and link  \textit{/sdcard/} to it.

\myparatight{File access}
An app can access a file using either a \emph{hard-coded approach} or a \emph{soft-coded approach}. Specifically, in the hard-coded approach, an app specifies an absolute file path (e.g., \textit{/data/data/com.facebook.katana/files/a.txt}) and reads/writes to the file. In the soft-coded approach, the app uses an Android API to locate a file and then operates on the file. Table~\ref{table:fileAPI} shows some example Android APIs that can be used to find file paths. These APIs were found via reading through the latest version Android documentation and source code. 
Listing \ref{list:motivateExample} shows an example of soft-coded file access. The app uses an API \textit{openFileOutput(fileName, 0)} at line 4 to open the file \textit{/data/data/<package name>/files/locSink}, and then the app writes GPS location and time to the file.  

\subsection{Problem Definition}
Suppose a forensic investigator is investigating a crime and has collected a suspect's mobile device. 
Moreover, the forensic investigator has obtained a file system image from the device, e.g., through physical image extraction~\cite{androidImageMethod}. We note that it is a common practice in forensics that the forensic investigator can retrieve the file system image, e.g., using SRSRoot~\cite{SRSRoot}. 
There could be thousands of files on the device. {For example, we extracted the physical image of a Nexus 7 tablet that was used for around 5 years. The device has installed 90 apps (including both system and user apps), which generated  around 19K files.} The forensic investigator aims to identify the files on the device that could contain certain types of evidentiary data, e.g., GPS locations, visited URLs. We call the problem \emph{evidence identification problem}, and we formally describe it as follows:

\begin{definition}[Evidence Identification Problem]
Given an image of a device's file system and a type of evidentiary data, the evidence identification problem is to identify the files (if any) that contain the type of evidentiary data.  
\end{definition}

In this work, we focus on solving the evidence identification problem for Android smartphones. Moreover, we focus on the evidentiary data including  \emph{location}, \emph{time}, \emph{visited URL}, and \emph{text input} as they were shown to be useful digital evidence in real-world crime investigations~\cite{WebSearchNews,uiucCaseComplaint,gpsCaseShoot,gpsCasePoach}. 
For instance, GPS locations and texts retrieved from suspects' smartphones were used to corroborate charges in one of the biggest poaching cases in Washington~\cite{gpsCasePoach}; visited URLs retrieved from a suspect's smartphone are used as important digital evidence for investigating a kidnapping case that happened at University of Illinois Urbana-Champaign in 2017. 
Although we focus on these types of evidentiary data, our techniques can be easily extended to other types of data such as contacts, device ID, and sensors, if a forensic investigator is interested in analyzing such data for potential evidence.

\section{EviHunter}
\label{sec:analysisFramework}

\subsection{Overview}
\begin{figure}[!t]
	\centering
	\includegraphics[width=\linewidth]{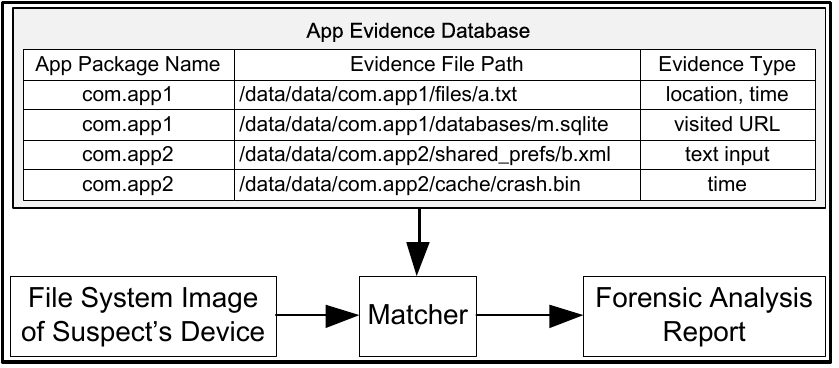}
	\caption{Overview of our EviHunter.}
	\label{fig:overview}
\end{figure}

Figure~\ref{fig:overview} overviews our EviHunter, which consists of two components, \emph{App Evidence Database} and \emph{Matcher}. 

\myparatight{App Evidence Database (AED)} AED contains the evidentiary data for a large number of apps.  
Specifically, each row of AED represents a file that could be generated by an app and the types of evidentiary data that the file contains. Files are identified by their \emph{file paths}.
AED has three columns: the first column includes apps' package names; the second column includes file paths; and the third column indicates the types of evidentiary data that the corresponding file could contain. 
 A file path could be either \emph{static} or \emph{dynamic}. Specifically, if a file path does not depend on the execution environment of the app that generates the file, then the file path is static, otherwise it is dynamic. Therefore, if an app uses a static file path for a file, then this file has the same path on different devices; if an app uses a dynamic file path for a file, then the file path could be different on different devices and at different times. For instance, an app could use \emph{timestamp} as a part of a file's name, which results in a dynamic file path. In our AED, a dynamic file path includes the pattern for the dynamic part. For instance, \textit{/data/data/<package name>/files/evidence-<timestamp>.txt} represents that the app could generate a file whose file name includes the timestamp when the file was generated.  

\myparatight{Matcher} Given a file system image from a suspect's device, the Matcher matches the file paths on the device against those in the AED, to identify the files on the device that could contain the evidentiary data of interest to forensic investigators. If the device has an app that is not included in AED, we will analyze the app using EviHunter and add the results into AED. 
A file path on the device is matched against a file path in the AED if they are the same. 

Moreover, when matching a file path against a dynamic file path in the AED, a forensic investigator can use partial match via ignoring the dynamic part or consider the pattern (e.g., date format) of the dynamic part via regular expression.

Next, we discuss how EviHunter builds the AED.

\begin{table*}[]

\centering

\begin{threeparttable}

\caption{Propagation rules of non-method-invoking statements. $C_a$ is a constant string, while $C_b$ is a constant that is not string. $f_A$ and $f_B$ are static fields, while $f_a$ and $f_b$ are instance fields. }
\label{table:propagateRule}

\def\arraystretch{1.2}

\begin{tabular}{lll}

\hline

\multirow{2}{*}{\textbf{Statement}} & \multicolumn{2}{c}{\textbf{Propagation Rule}} \\ \cline{2-3}

& \textbf{\textit{EvSet}} & \textbf{\textit{Path}}\\ \hline

$v_a = C_a$& $t(v_a).EvSet \leftarrow \emptyset$ & $t(v_a).Path \leftarrow C$ \\
$v_a = C_b$& $t(v_a).EvSet \leftarrow \emptyset$ & $t(v_a).Path \leftarrow \emptyset$ \\

$v_a = v_b$ & $t(v_a).EvSet \leftarrow t(v_b).EvSet$ & $t(v_a).Path \leftarrow t(v_b).Path$ \\

$v_a = f_B$ & $t(v_a).EvSet \leftarrow t(f_B).EvSet$ & $t(v_a).Path \leftarrow t(f_B).Path$ \\

$f_A = v_b$ & $t(f_A).EvSet \leftarrow t(v_b).EvSet$ & $t(f_A).Path \leftarrow t(v_b).Path$ \\

$v_a = v_b.f_b$ & $t(v_a).EvSet \leftarrow t(v_b.f).EvSet$ & $t(v_a).Path \leftarrow t(v_b.f).Path$ \\

$v_a.f_a = v_b$ & $t(v_a.f).EvSet \leftarrow t(v_b).EvSet$ & $t(v_a.f).Path \leftarrow t(v_b).Path$ \\

$v_a = v_b[v_c]$ & $t(v_a).EvSet \leftarrow t(v_b).EvSet$ & $t(v_a).Path \leftarrow t(v_b).Path$ \\

$v_a[v_b] = v_c$ & $t(v_a).EvSet \leftarrow t(v_a).EvSet \ \cup \ t(v_c).EvSet$ & $t(v_a).Path \leftarrow t(v_c).Path$ \\

$v_a = v_b \ + \ v_c$\tnote{1} & $t(v_a).EvSet \leftarrow t(v_b).EvSet\ \cup\ t(v_c).EvSet$ & $t(v_a).Path \leftarrow t(v_b).Path + t(v_c).Path$ \\

$v_a = v_b \ \textit{binop} \ v_c$\tnote{2} & $t(v_a).EvSet \leftarrow t(v_b).EvSet\ \cup\ t(v_c).EvSet$ & $t(v_a).Path \leftarrow \emptyset$ \\ \hline

\end{tabular}

\begin{tablenotes}
\item [1] When $v_b$ and $v_c$ are string values.
\item [2] When the binary operator \textit{binop}  is not ``+''. 
\end{tablenotes}
\end{threeparttable}

\end{table*}

\subsection{Building the AED via Static Analysis}
AED should have a high coverage to avoid missing potential evidentiary data. Specifically, for an app, AED should include as many file paths that contain evidentiary data as possible.
we leverage static analysis instead of dynamic analysis to build the AED. In particular, we develop a static {data-flow analysis} method to build an AED for a large number of apps. Specifically,  in our static data-flow analysis, we define a customized tag structure for variables and propagate the tags in an app via \emph{forward analysis}. We leverage existing sink methods for file system found by existing tools~\cite{flowdroid,susi,droidsafe}. However, we uncover new sources as the sources in existing tools are not complete.

\subsubsection{Pre-processing}
Given an app, we first use Soot~\cite{soot} to transform the app to Jimple code, a three address
 intermediate representation. Second, we use IC3~\cite{ic3} to build the inter-connected component communication (ICC) models. Finally, we use FlowDroid~\cite{flowdroid} to construct call graphs and entry points.  The latest version of FlowDroid has integrated IccTA~\cite{iccta} to incorporate the ICC models when building call graphs. 
FlowDroid also extracts the app's package name, which we will use as the app's identifier.
We note that FlowDroid alone is insufficient to build our AED. Specifically, FlowDroid aims to identify data flows from sources to sinks. 
However, FlowDroid does not identify the file paths where data are written to.

\subsubsection{Tag for a Variable}
We define a tag structure for each variable, e.g., primitive, object, and class field. Then, we propagate variables' tags via performing forward data-flow analysis starting from the entry point in the call graph. We denote by $t(v)$ the tag for a variable $v$.
The tag should carry sufficient information to identify both the types of evidentiary data and the file paths. 
To achieve these goals, we propose a tag structure which includes the following information: 
\begin{itemize}
\item \textit{Evidence type set {(EvSet)}}: Types of evidentiary data that a variable could contain.
\item \textit{File path {(Path)}}:  File path associated with a variable. 
For instance, when an app writes data to file system via a file descriptor, the \textit{Path} associated with the file descriptor is the file where data are written to. 
\end{itemize}

We note that in conventional static analysis tools that were designed to detect sensitive data flows in Android apps, a tag often only includes the data types (i.e., \textit{EvSet} in our case).

\subsubsection{Propagation Rules}
\begin{table*}[!h]

\caption{Data-flow summary of example Android APIs for file access.}
\label{table:getdirectory}
\centering
\def\arraystretch{1.2}
\makebox[\linewidth]{\small

\begin{threeparttable}

\begin{tabular}{|c|l|}
\hline
\multicolumn{1}{|c|}{Method} & \multicolumn{1}{c|}{Data-flow Summary} \\ \hline

\textit{getDataDirectory()} & $t(v_0).\textit{Path} \leftarrow$ "/data/" \\ \hline

\textit{getCacheDir()} & $t(v_0).\textit{Path} \leftarrow$ "/data/data/\textless package name\textgreater /cache/"  \\ \hline

\textit{getFilesDir()} & $t(v_0).\textit{Path} \leftarrow$ "/data/data/\textless package name\textgreater/files/" \\ \hline

\textit{getFileStreamPath(String $v_2$)} & $t(v_0).\textit{Path} \leftarrow$ "/data/data/\textless package name\textgreater/files/" + $v_2$ \\ \hline

\textit{openFileOutput(String $v_2$, int $v_3$)} & $t(v_0).\textit{Path} \leftarrow$ "/data/data/\textless package name\textgreater/files/" + $v_2$ \\ \hline

\textit{getDataDir()} & $t(v_0).\textit{Path} \leftarrow$ "/data/data/\textless package name\textgreater/databases/" \\ \hline

\textit{openOrCreateDatabase(String $v_2$, ... )} & $t(v_0).\textit{Path} \leftarrow$ "/data/data/\textless package name\textgreater/databases/" + $v_2$ \\ \hline

\textit{openDatabase(String $v_2$, ... )} & $t(v_0).\textit{Path} \leftarrow$ "/data/data/\textless package name\textgreater/databases/" + $v_2$ \\ \hline

\textit{getSharedPreferences(String $v_2$,int $v_3$)} & $t(v_0).\textit{Path} \leftarrow$ "/data/data/\textless package name\textgreater/shared\_prefs/" + $t(v_1).\textit{Path}$ + ".xml" \\ \hline

\textit{getDefaultSharedPreferences(Context $v_2$)} & $t(v_0).\textit{Path} \leftarrow$ "/data/data/\textless package name\textgreater/shared\_prefs/\textless package name\textgreater\_preferences.xml" \\ \hline

\textit{getPreferences(int $v_2$)} & $t(v_0).\textit{Path} \leftarrow$ "/data/data/\textless package name\textgreater/shared\_prefs/\textless context\textgreater.xml" \\ \hline

\textit{getDir(String $v_2$, int $v_3$)} & $t(v_0).\textit{Path} \leftarrow$ "/data/data/\textless package name\textgreater/app\_" + $v_2$ \\ \hline

\textit{openOrCreateDatabase(File $v_2$, ... )} & $t(v_0).\textit{Path} \leftarrow$ $t(v_2).Path$ \\ \hline


\textit{getExternalStorageDirectory()} & $t(v_0).\textit{Path} \leftarrow$ "/sdcard/" \\ \hline

\textit{getExternalStoragePublicDirectory(String $v_2$)} & 
\begin{tabular}[c]{@{}l@{}}
$t(v_0).\textit{Path} \leftarrow$ "/sdcard/", if $v_2$ is empty.\\ 
$t(v_0).\textit{Path} \leftarrow$ "/sdcard/" + $v_2$ + "/", otherwise.
\end{tabular} \\ \hline

\textit{getObbDir()} & $t(v_0).\textit{Path} \leftarrow$ "/sdcard/Android/obb/\textless package name\textgreater/" \\ \hline

\textit{getExternalCacheDir()} & $t(v_0).\textit{Path} \leftarrow$ "/sdcard/Android/data/\textless package name\textgreater/cache/" \\ \hline

\textit{getExternalFilesDir(String $v_2$)} & 
\begin{tabular}[c]{@{}l@{}}
$t(v_0).\textit{Path} \leftarrow$ "/sdcard/Android/data/\textless package name\textgreater/files/", if $v_2$ is empty.\\ 
$t(v_0).\textit{Path} \leftarrow$ "/sdcard/Android/data/\textless package name\textgreater/files/" + $v_2$ + "/", otherwise.
\end{tabular} \\ \hline

\end{tabular}



\end{threeparttable}}

\end{table*}

Propagation rules define how tags are updated when analyzing the statements in an app. Our rules are applied to the three-address Jimple code of an app. 
We classify statements into two groups, i.e., \emph{non-method-invoking statements} and \emph{method-invoking statements}.
We discuss the propagation rules for them separately.

\myparatight{Non-method-invoking statements}  Table \ref{table:propagateRule} shows the propagation rules for the possible non-method-invoking statements in the Jimple intermediate representation. When a statement assigns a constant string to a variable, we set the \textit{EvSet} and  \textit{Path} of the variable's tag to be empty and the constant, respectively; if the constant is not a string, we will set both  \textit{EvSet} and  \textit{Path} to be empty. Moreover, when a statement assigns one variable's value to another variable, i.e., $v_a=v_b$, we assign $v_b$'s tag to $v_a$. Next, we discuss the statements with more complex assignment relationships.   

\begin{itemize}
\item \emph{Class field access}: 
A class field can be a static field or an instance field. A static field is shared by all instances of a class, while an instance field is unique to an instance of a class. 
Each field has a tag. We maintain static fields of a class in a globally available map, where static fields are keys and their tags are values in the map. 
For an instance, we use a map to maintain its instance fields and their tags, where fields are keys and their tags are values in the map. This map structure can help track the fields of an object.

\item \emph{Array access}: For an array variable $v_a$, we store its tag $t(v_a)$ as the union of the tags of its elements. Moreover, for each array, we maintain a map that stores the tags of its elements. For instance, $t(v_a.i)$ is the tag of the $i$th element of the array variable $v_a$.  For each array access, we resolve the index whenever we can.  Suppose we have an assignment statement $v_a[v_b] = v_c$. If the index variable $v_b$  can be resolved, we propagate tags as $t(v_a.v_b) \leftarrow t(v_c)$, otherwise we merge the evidence set $t(v_c).EvSet$ of the variable $v_c$ into that of $v_a$. In the case of the statement $v_a = v_b[v_c]$, if $v_c$ is resolvable, we assign the tag of the element $v_c$ to $v_a$, otherwise we assign the tag of the array variable $v_b$ to that of the variable $v_a$. The propagation rules for array access shown in Table \ref{table:propagateRule} are for the scenarios where index cannot be resolved.

\item \emph{Binary operator}: We propagate the union of the \emph{EvSet} of the two operands for all binary operators.  
If the operator is $+$ and both operands are string, then we update the \emph{Path} as the concatenation of the \emph{Path} of the two operands, otherwise we set the \emph{Path}  to be empty. 
\end{itemize}

\begin{align}
\label{eq:callFlowExample}
v_{1}.method(v_{2}, v_{3}, \cdots); \\
\label{eq:returnFlowExample}
v_{0} = v_{1}.method(v_{2}, v_{3}, \cdots);
\end{align}

\myparatight{Method-invoking statements} A method-invoking statement could be a method call without leveraging the return value or a method call with assigning the return value to a variable. We abstract the two cases in Equation~\ref{eq:callFlowExample} and \ref{eq:returnFlowExample}, respectively. In the first case, we propagate the tags of the base instance $v_{1}$ (if available) and arguments $v_{2}, v_{3}, \cdots $ into the callee and analyze the data flow in the callee. In the second case, we will analyze data flow in the callee and assign the tag of the return value to that of the variable $v_{0}$. We note that when analyzing method calls, we may get into a loop of method calls. For example, if there is a loop in the call graph of an app. We avoid the loop by using a stack to keep track of method calls and skipping a method call if the method is already on the stack. Essentially, we analyze the methods in a loop in the call graph once.


\subsubsection{Multi-threading and Reflection} 
Multi-threading and reflection are well known challenges for static program analysis. We leverage the techniques from the state-of-the-art static analysis tools~\cite{droidsafe,HornDroid,rDroid} to partially handle  multi-threading and reflection. Specifically, for multi-threading, we assume the threads execute in a sequential order, following prior studies~\cite{HornDroid,rDroid}. Therefore, whenever a certain thread is spawned and starts running, we find its corresponding entry method and redirect analysis to it.
For example, an instance invoke of method \texttt{start()} in \texttt{java.lang.Thread} will be redirected to its actual running method \texttt{run()}.
We handle the dedicated Android threading library \texttt{android.os.AsyncTask} and \texttt{android.os.Handler} by method redirecting as well.
For reflection, we analyze the reflective call only if the method can be parsed statically, following prior static analysis tools~\cite{droidsafe,HornDroid}.
In particular, EviHunter uses the parsed declaring class name and \textit{Path} information to determine the actual method call and redirect the reflective call to it.

\subsubsection{Data-Flow Summary for System APIs}
For complexity consideration, we manually summarize the data flows for the commonly used system APIs and skip the remaining ones,  instead of incorporating the framework code into EviHunter. We note that DroidSafe~\cite{droidsafe} proposed a technique to model framework, which was able to find more sensitive data flows in benchmark apps. However, Reaves \emph{et al.}~\cite{starDroid}  found that DroidSafe requires a large amount of main memory and fails to analyze real-world apps. In addition, DroidSafe does not track the file paths where data are written to. Therefore, we do not adopt the  technique in DroidSafe to model framework.

When we discuss the data-flow summary for system APIs, we will refer to Equation~\ref{eq:callFlowExample} and~\ref{eq:returnFlowExample} for the definition of variables.  

\begin{table}[]
\centering

\caption{Data-flow summary of example Java APIs for file path construction.}
\label{table:fileconstructor}
\bgroup
\def\arraystretch{1.2}

\makebox[\linewidth]{

\begin{tabular}{|c|l|c|}
\hline
\multicolumn{1}{|c|}{Method} & \multicolumn{1}{c|}{Data-Flow Summary} \\ \hline

\textit{FileWriter \textless init\textgreater(File $v_2$)} & $t(v_1).\textit{Path}\ \leftarrow\ t(v_2).Path$ \\ \hline

\textit{File \textless init\textgreater(File $v_2$, String $v_3$)} & $t(v_1).\textit{Path}\ \leftarrow\ t(v_2).Path\ +\ v_3$\\ \hline

\end{tabular}
}
\egroup

\end{table}

\myparatight{System APIs to construct file paths} Java provides APIs for apps to construct file paths and access files in a hard-coded approach. We summarize data flows for these APIs. Table~\ref{table:getdirectory} shows the data-flow summary for some example Android APIs that are used to get file paths. Table~\ref{table:fileconstructor} shows data-flow summary for example APIs.  
 As we described in Section~\ref{sec:filesystem}, Android provides APIs for apps to locate files in a soft-coded approach. 
A majority of these Android APIs can be summarized using their arguments and the app's package name. For instance,  the API \textit{getExternalFilesDir(String)} returns a file object whose file path is 
\textit{/sdcard/Android/data/\textless package name\textgreater/files/} if the input parameter is an empty string, otherwise the file path is \textit{/sdcard/Android/data/\textless package name\textgreater/files/$v_2$/}, where $v_2$ is the input parameter.

However, there are two cases that require extra information to summarize the file paths for Android APIs.
The first one is \textit{getPreferences(int)} that returns a SharedPreferences file whose file path depends on the Context class.
Context refers to the runtime environment, for example, a launching Activity named "MainActivity".
If "MainActivity" creates a SharedPreferences file by \textit{getPreferences(int)}, the file name is "MainActivity.xml".
To handle this case, we trace back the method calls until finding the Context and we use the class name of the Context to resolve the corresponding SharedPreferences file path.

The second case is the SQLite database creation and access through \textit{SQLiteOpenHelper}. 
An app can access a SQLite database via creating a class inheriting this helper class and handling the inherited callback methods.
When the inherited class is initialized, a SQLite database is created through the helper class initialization method \textit{SQLiteOpenHelper\textless init\textgreater(Context, String, $\cdots$)}, where the $2^{nd}$ argument is the database name and the database is located in the directory \textit{/data/data/\textless package name \textgreater/databases/}. When \textit{getWritableDatabase()} is called to retrieve database object, we search the declaring class name and assign the database file path to the corresponding variable's tag.

We note that these Android  APIs have changed over Android versions. Therefore, in practice, our tool requires summarizing data flows for these Android APIs for different versions.

\myparatight{System APIs for string operations and commonly used data structures} W summarize the data flows of APIs for string operations. Example APIs include \textit{toString()}, \textit{valueOf()}, \textit{<init>(String)} (for string initialization), and \textit{concat(String)}. Moreover, we summarize the data flows for collection class, string buffers, and similar commonly used data structures such as HashSet and ArrayList. Table~\ref{table:tagpropagation} shows our data-flow summary for some example APIs.
\begin{table}[]
\centering
\caption{Data-flow summary of example APIs for string operation and commonly used data structures.}
\label{table:tagpropagation}
\bgroup
\def\arraystretch{1.2}

\makebox[\linewidth]{

\begin{tabular}{|c|l|}
\hline
  API  & \multicolumn{1}{c|}{Data-Flow Summary} \\ \hline

\textit{toString()} & $t(v_0) \leftarrow t(v_1)$ \\ \hline

 \textit{valueOf(double $v_2$)} & $t(v_0) \leftarrow t(v_2)$ \\ \hline

\textit{read(byte[] $v_2$)} & $t(v_1) \leftarrow t(v_2)$ \\ \hline

\textit{\textless init\textgreater(String $v_2$)} & 
\begin{tabular}[l]{@{}l@{}}
$t(v_0).\textit{Path} \leftarrow v_2$\\ 
$t(v_0).\textit{EvSet} \leftarrow t(v_2).\textit{EvSet}$ 
\end{tabular}
\\ \hline

\textit{concat(String $v_2$)} & 
\begin{tabular}[l]{@{}l@{}}
$t(v_0).\textit{Path} \leftarrow t(v_1).\textit{Path} + v_2$\\ 
$t(v_0).\textit{EvSet} \leftarrow t(v_2).\textit{EvSet} \cup t(v_1).\textit{EvSet}$ 
\end{tabular}

\\ \hline

\textit{add(Object $v_2$)} & $t(v_1).\textit{EvSet} \leftarrow t(v_1).\textit{EvSet} \cup t(v_2).\textit{EvSet}$ \\ \hline

\end{tabular}%
}

\egroup

\end{table}


\begin{lstlisting}[float, frame=single,numbersep=-10pt,numbers=left,language=java,caption={Method arguments as sources for URL.},label=list:browsingHistorySameple]
    void foo(){
        WebView wv = 
            (WebView)findViewById(R.id.webView);
        String testUrl = "http://www.foo.com";    
        wv.loadUrl(testUrl);
        sink(testUrl);
        wv.setWebViewClient(new WebViewClient(){            
            @Override
            public void onPageFinished
            (WebView view, String url){
                sink(url);
            }
        });    
    }
\end{lstlisting}

\myparatight{System native methods} We obtain a list of system native methods from DroidSafe~\cite{droidsafe}. For each system native method call,  we make an over-approximation of evidence types for each input/output variable involved in the method call. Specifically, we take the union of the evidence set \textit{EvSet} of the input variables, and we assign the union to each input variable and output variable. We update input variables because they may be modified within the native method. 

 
\begin{table*}[]
\caption{The AED that EviHunter constructed for the 4 benchmark apps in DroidBench.  <internal storage> refers to ``/data/data/<package name>''.}
\label{table:droidBenchResult}

\centering
\def\arraystretch{1.2}
\makebox[\linewidth]{%


\begin{tabular}{|c|l|l|}

\hline
\textbf{App Package Name} & \multicolumn{1}{c|}{\textbf{Evidence File Path}} & \multicolumn{1}{c|}{\textbf{Evidence Type}} \\ \hline

 de.ecspride & \textless internal storage\textgreater /files/out.txt & Device ID \\ \hline 

 edu.mit.event\_context\_shared\_pref\_listener & \textless internal storage\textgreater /shared\_prefs/settings.xml & Device ID \\ \hline 
 
 edu.mit.icc\_event\_ordering & \textless internal storage\textgreater /shared\_prefs/prefs.xml & Device ID \\ \hline

 org.cert.writeFile & \textless internal storage\textgreater /files/sinkFile.txt & Location \\ \hline

\end{tabular}}%

\end{table*}

\begin{table*}[]
\caption{The AED that EviHunter constructed for the three benchmark apps we designed.  <internal storage> refers to ``/data/data/<package name>'',  while <external storage> represents ``/sdcard/Android/data/<package name>''.}
\label{table:fileBenchResult}

\centering
\def\arraystretch{1.2}
\makebox[\linewidth]{%

\begin{threeparttable}[]

\addtolength{\tabcolsep}{-4pt}

\begin{tabular}{|c|l|l|}

\hline
\textbf{App Package Name} & \multicolumn{1}{c|}{\textbf{Evidence File Path}} & \textbf{Evidence Type} \\ \hline

{com.evihunter.GPS} & \textless internal storage\textgreater /shared\_prefs/com.evihunter.GPS\_preferences.xml & Time \\ \hline 

{com.evihunter.GPS}& \textless internal storage\textgreater /databases/mfGps.db & Location, Time \\ \hline 
 
{com.evihunter.GPS} & \textless internal storage\textgreater /files/\textless timestamp\textgreater.txt & Location \\ \hline


com.evihunter.Browser & \textless internal storage\textgreater /app\_goo2/goo3 & Visited URL \\ \hline 
 
 com.evihunter.Browser& \textless internal storage\textgreater /files/foo\_\textless UUID\textgreater.bin & Time \\ \hline
 
com.evihunter.Browser & \textless external storage\textgreater /files/foo3 & Visited URL \\ \hline 
 
  
com.evihunter.Browser & /sdcard/browser\_\textless intent\textgreater.txt & Time \\ \hline
 
com.evihunter.IM & \textless internal storage\textgreater/databases/mfChat.db & Text Input, Time \\ \hline
 

\end{tabular}%



\end{threeparttable}}

\end{table*}

\subsubsection{Sources and Sinks}
\label{sec:subsubSrcSink}

In EviHunter, a sink is a system API that writes data to file system, while a source is where evidentiary data are created or file path is created. 
We first combined the publicly available sources and sinks in existing tools including FlowDroid~\cite{flowdroid}, SuSi~\cite{susi}, and DroidSafe~\cite{droidsafe}. These sources and sinks were also used by more recent tools, e.g., HornDroid~\cite{HornDroid}.

We use the sink methods combined from existing tools. However, we found that the combined source methods  for evidentiary data are not complete. In particular, existing tools missed the methods whose arguments indicate sources. Moreover, they did not consider sources for file paths. Therefore, we extend the source methods for evidentiary data and uncover the sources for file paths. We make the sources and sinks used by EviHunter publicly available~\cite{sourcesink}.

\myparatight{Sources for evidentiary data (\emph{EvSet})} In this work, we focus on the types of evidentiary data including \emph{location}, \emph{text input}, \emph{time}, and \emph{visited URL} as they were shown to be useful in real-world crime investigations~\cite{WebSearchNews,uiucCaseComplaint,gpsCaseShoot,gpsCasePoach}. However, our EviHunter can be extended to other types of evidentiary data if needed. Specifically, we can add the sources for those types of evidentiary data in EviHunter and extend the \textit{EvSet} in tags. Next, we discuss the sources for each of the four types of evidentiary data.  

{\bf 1) Location:} Location includes GPS location and course-grained location determined by WiFi and/or cellular data. We treat an Android API that returns location data as a source.
We obtained 39 source methods for location from existing tools~\cite{flowdroid,susi,droidsafe}. Moreover, we found that GPS location can also be created in the argument of a method. Specifically, the argument of the method \emph{onLocationChanged(android.location.Location)} stores a GPS location. Therefore, when we analyze a statement that involves \emph{onLocationChanged}, we will add the data type \emph{location} to the argument's tag.   

{\bf 2) Text input:} Text input is the string data typed in by users. For example, an instant message in a social networking app is a text input; a search query is a text input. We obtained 2 source methods for text input from existing tools~\cite{flowdroid,susi,droidsafe}.

{\bf 3) Time:} We obtained 16 source methods for time from existing tools~\cite{flowdroid,susi,droidsafe}. Moreover, we found one more source method for time, i.e., \textit{currentTimeMillis()}, which returns the system timestamp.

{\bf 4) Visited URL:} A user could visit URLs via a browser or a non-browser app using WebView. 
We obtained 3 source methods that return URLs from existing tools~\cite{flowdroid,susi,droidsafe}, and we found one more source method whose return value is URL.
Moreover, we found that the arguments of certain Android APIs and callback methods correspond to visited URLs. Listing \ref{list:browsingHistorySameple} shows an example. 
\textit{testUrl} is initialized as a string constant, passed to the Android API \textit{loadUrl(testUrl)} as an input argument, and finally written to the file system.  
We can identify that the \textit{testUrl} is a visited URL because it is used as an argument of  the Android API \textit{loadUrl(String)}. Therefore, once a variable is used as an argument of \textit{loadUrl(String)}, we add the data type \emph{visited URL} to the \emph{EvSet} of the variable's tag.

Arguments of certain callback methods also indicate URLs. For instance, the second argument of the callback method  \textit{onPageFinished(...)} (e.g., line 9 in Listing \ref{list:browsingHistorySameple}) corresponds to a URL. In total, we find 6 callback methods whose arguments correspond to URLs. Once a variable is passed as the corresponding argument of these methods, we extend the variable's tag to include \emph{visited URL}. 

\myparatight{Sources for file paths (\emph{Path})} An app could use a static file path or a dynamic file path. When a constant string is assigned to a variable, we initialize the variable's \emph{Path} as the constant. 
In order to understand dynamic file paths, we sampled 100 dynamic file paths in our preliminary analysis results and did a manual measurement study about them. Via manually analyzing the code, we found the top-3 ways that apps use to generate dynamic file paths include \emph{intent},  \emph{timestamp}, and  \emph{universally unique identifier (UUID)}, which represent 33\%, 20\%, and 12\% of the dynamic file paths, respectively. Intent refers to the case where a part of the file path is constructed from an intent that is used for  inter-component communications. When a variable is assigned as the return value of a method that corresponds to intent, we assign the variable's \emph{Path} as \emph{<intent>}. We found 38 such methods.   
 UUID  is a 128-bit random string generated by the API \textit{randomUUID()} in the class \texttt{java.util.UUID}. When a variable is assigned as the return value of the API \textit{randomUUID()}, we initialize the variable's \emph{Path} as \emph{<UUID>}. Timestamp refers to the case where a part of the file path is constructed using the system time. When a variable is assigned as the return value of a system method that returns system time, we initialize the variable's \emph{Path} as \emph{<timestamp>}. We use the source methods for time that we discussed in the above as sources for \emph{<timestamp>}.

\begin{table*}[]
\caption{Analysis results for each type of evidentiary data on the 8,690 real-world apps. The column ``Static'' indicates the number of static file paths; the column ``Dynamic'' indicates the number of dynamic file paths; and the column ``App'' indicates the number of apps. ``Others'' indicates a file that does not include the four specified types of evidentiary data.}
\label{table:realworldEvidenceResult}

\centering
\def\arraystretch{1.2}
\makebox[\linewidth]{

\begin{threeparttable}

\begin{tabular}{|c|c|c|c|c|c|c|c|c|c|}
\hline
\multirow{2}{*}{\textbf{Evidence Type}} & \multicolumn{2}{c|}{\textbf{SQLite Database}} & \multirow{2}{*}{\textbf{App}} & \multicolumn{2}{c|}{\textbf{SharedPreferences}} & \multirow{2}{*}{\textbf{App}} & \multicolumn{2}{c|}{\textbf{Ordinary File}} & \multirow{2}{*}{\textbf{App}} \\ \cline{2-3} \cline{5-6} \cline{8-9}
                                        & \textbf{Static}       & \textbf{Dynamic}      &                               & \textbf{Static}        & \textbf{Dynamic}       &                               & \textbf{Static}  & \textbf{Dynamic}  &                               \\ \hline
\textbf{Location}                       & 145                    & 0                     & 145                            & 195                    & 0                      & 72                            & 151              & 0                 & 151                           \\ \hline
\textbf{Time}                           & 343                   & 0                     & 316                           & 1128                   & 1                      & 924                           & 431              & 4                 & 411                           \\ \hline
\textbf{Visited URL}                    & 20                    & 0                     & 19                            & 25                     & 0                      & 25                            & 19               & 0                 & 18                            \\ \hline
\textbf{Text Input}                     & 166                   & 0                     & 155                           & 410                    & 1                      & 388                           & 184              & 2                 & 148                           \\ \hline
\textbf{Others}                        & 903                   & 3                     & 721                           & 5941                   & 3                      & 3135                          & 6901             & 576               & 3448                          \\ \hline
\end{tabular}

\end{threeparttable}}

\end{table*}

\begin{table}[]
\centering
\caption{Summary of the analysis results for the 8,690 real-world apps.}
\label{table:realworldPathResult}
\vspace{-2mm}
\centering
\def\arraystretch{1.2}
\makebox[\linewidth]{

\begin{tabular}{ccc}
\hline
\multicolumn{1}{|c|}{\multirow{2}{*}{\textbf{File Type}}} & \multicolumn{2}{c|}{\textbf{Evidence File Path}} \\ \cline{2-3} 
\multicolumn{1}{|c|}{}                            & \multicolumn{1}{c|}{\textbf{Static File Path}}                    & \multicolumn{1}{c|}{\textbf{Dynamic File Path}} \\ \hline

\multicolumn{1}{|c|}{\textbf{SQLite Database}}             & \multicolumn{1}{c|}{674}        & \multicolumn{1}{c|}{0}     \\ \hline

\multicolumn{1}{|c|}{\textbf{SharedPreferences}}          & \multicolumn{1}{c|}{1758}        & \multicolumn{1}{c|}{2}       \\ \hline

\multicolumn{1}{|c|}{\textbf{Ordinary File}}                        & \multicolumn{1}{c|}{785}      & \multicolumn{1}{c|}{6}    \\ \hline

\multicolumn{1}{|c|}{\textbf{Total}}          & \multicolumn{1}{c|}{3217}        & \multicolumn{1}{c|}{8}       \\ \hline
\end{tabular}}
\vspace*{-0.3cm}
\end{table}




\section{Evaluation}
\label{sec:evaluation}

We aim to evaluate the AED generated by EviHunter. First, we evaluate EviHunter using benchmark apps from DroidBench and benchmark apps that we design. 
Second,  we evaluate EviHunter on a large number of real-world apps. 
Third, we show a case study on how a forensic investigator can use EviHunter to find evidentiary data on an Android smartphone. 

Our implementation leverages Soot~\cite{soot} to transform an app to Jimple code, IC3~\cite{ic3} to build inter-component communications models, and FlowDroid~\cite{flowdroid} to build call graphs and entry points. We perform our experiments on an Intel\textsuperscript{\textregistered} Xeon\textsuperscript{\textregistered} CPU E5-1603 v3 @ 2.8GHz running Ubuntu 14.04 with 64GB of heap memory for the JVM.

\subsection{Results on Benchmark Apps}

Previous studies on mobile security and privacy have designed and published some benchmark apps, e.g., DroidBench~\cite{droidBench} provides a collection of benchmark apps. 
Among the 120 benchmark apps in DroidBench, we found only 4 apps have sinks to a file system and they all use the soft-coded approach to construct file paths.  One possible reason is that these benchmark apps were designed for privacy studies and storing sensitive data only at the local file system without sending them to the Internet may not be considered as a privacy leakage. Among the 4 apps, 3 of them consider device ID as sensitive data and the remaining one considers GPS location as the sensitive data. To analyze these benchmark apps, we extend EviHunter to incorporate device ID as evidentiary data. Specifically, we add the system APIs that return device ID to the sources. 
Table~\ref{table:droidBenchResult} shows the AED that EviHunter constructed for the 4 benchmark apps. EviHunter finds all the file paths accurately in the 4 apps.
We note that in the app org.cert.WriteFile, the GPS location data is passed between components. 
EviHunter accurately identifies the location data. This is because EviHunter leverages the inter-component communications support in FlowDroid to track evidentiary data across components. 

However, the 4 benchmark apps have limitations at evaluating EviHunter. Specifically, they all use the soft-coded approach to construct file paths; they only consider SharedPreferences and ordinary files; and they only consider device ID and GPS location as data sources. To address these limitations and better evaluate EviHunter, we design three benchmark apps by ourselves.  
Specifically, our benchmark apps are \emph{GPS}, \emph{Browser}, and \emph{Instant Messenger}, whose package names are \emph{com.evihunter.GPS},  \emph{com.evihunter.Browser}, and \emph{com.evihunter.IM}, respectively. As their names suggest, these apps are designed to mainly evaluate evidentiary data \emph{location}, \emph{visited URL}, and \emph{text input}, respectively. The data type \emph{time} is covered by more than one benchmark app. When we design the benchmark apps, we consider  static file paths, dynamic file paths, soft-coded approach and hard-coded approach for file path construction, as we as SQLite database files, SharedPreferences files, and ordinary files. In particular, dynamic file paths include the three popular patterns <timestamp>, <UUID>, and <intent> that we discussed in Section~\ref{sec:subsubSrcSink}.

We use EviHunter to analyze the three benchmark apps. Table \ref{table:fileBenchResult} shows the AED EviHunter constructed for the three apps. EviHunter accurately finds all the file paths that could store evidentiary data for the three benchmark apps.   

\subsection{Results on Real-World Apps}
\label{sec:evaluation_d_realworldResult}

We obtained 8,690 real-world Google Play apps collected by PlayDrone~\cite{PlayDrone}. We use EviHunter to build an AED for these apps. 
Some apps take a long time to be fully analyzed. Since we aim to analyze a large number of apps,  we set a 3-minute timeout for each real-world app analysis. Specifically, if our tool does not finish analyzing an app within 3 minutes, we force the analysis to abort and report the analysis results. Note that the 3 minutes timeout does not count the time used for preprocessing an app via Soot, IC3, and FlowDroid. 
Our analysis stops early for 583 apps (6.7\% of total apps) due to the 3-minute timeout. 

Table \ref{table:realworldPathResult} summarizes our analysis results, and Table \ref{table:realworldEvidenceResult} further shows the analysis results for each type of evidentiary data. A reported file could include at least one type of evidentiary data including location, visited URL, time, and/or text input. A file path is treated as dynamic file path if the file path includes one of the three patterns <timestamp>, <UUID>, and <intent>. All other paths are treated as static file paths. Our manual measurement study (discussed in Section~\ref{sec:subsubSrcSink}) showed that around 65\% of dynamic file paths use the patterns <timestamp>, <UUID>, and/or <intent>. In other words, around 35\% of dynamic file paths are treated as static file paths. Therefore, a small number of the static file paths shown in the table are actually dynamic file paths. 

First, static file paths are much more frequently used than dynamic file paths by app developers. Specifically, since around 65\% of dynamic file paths use the patterns <timestamp>, <UUID>, and/or <intent>, the total number of dynamic file paths is around 12. Therefore, around 0.4\% of file paths are dynamic file paths. Second, SharedPreferences are more frequently used in apps than SQLite database and ordinary files (e.g., text and binary files). The reason may be that SharedPreferences are well defined lightweight data structures and provide rich APIs for developers to maintain data easily. Interestingly, in dynamic file paths, ordinary files are more frequently used than SQLite database and SharedPreferences.

\begin{figure}[!h]
\centering
\includegraphics[width=0.45\textwidth]{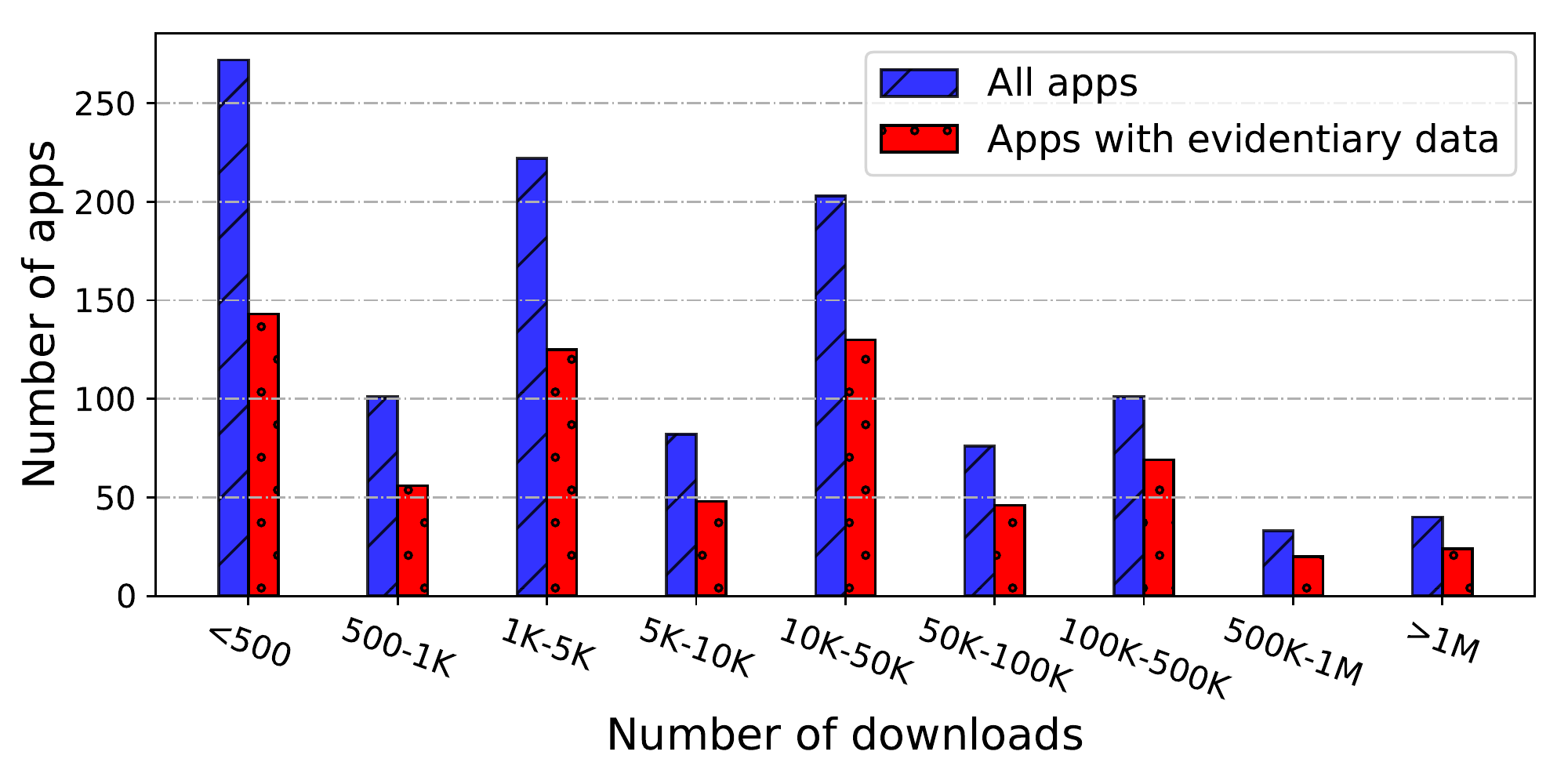}
\caption{App popularity.}
\label{fig:popularity}
\end{figure}

\myparatight{App popularity} We randomly sampled some apps and collected their metadata on Google Play. In total, we collected metadata of 1,130 randomly sampled apps. Among these apps,  EviHunter reports that 661 of them store evidentiary data in the file system. Figure~\ref{fig:popularity} shows the number of apps that have a given number of downloads. We observe that the apps that store evidentiary data on the file system have diverse popularity, ranging from hundreds of downloads to millions of downloads. Moreover, whether an app stores evidentiary data on the file system does not depend on its popularity. Specifically, around half of the apps in each category of popularity (e.g., $<500$, 5K-10K, or $>$1M) store evidentiary data on the file system.

\myparatight{Manual verification} Without ground truth, it is challenging to evaluate our results for the real-world apps. We perform a best-efforts manual verification. Specifically, we randomly sampled 30 apps that EviHunter reports to have at least one file containing evidentiary data.  We installed each app on a smartphone, clicked as many buttons of the app as possible, and typed in text inputs when we can. Then, we manually analyzed each file generated by the apps. In total, these apps generated 559 files. EviHunter reported that 72 of them could contain evidentiary data. 

For a given type of evidentiary data (e.g., GPS), a file is a false positive if EviHunter reports that the file  includes the evidentiary data but the file actually does not; and a file is a false negative if EviHunter  reports that the file does not  include the evidentiary data but the file actually does. We note that a file could be both false positive and false negative with respect to different types of evidentiary data. For instance, suppose a file includes GPS data and EviHunter reports that the file includes visited URL. Therefore, this file is a false positive with respect to visited URL and a false negative with respect to GPS. We compute the precision and recall for each type of evidentiary data.  We find that  EviHunter achieves a precision of 90\% and a recall of  89\% averaged over the four types of evidentiary data considered in the paper. 
 Additionally, we randomly sampled 30 apps that EviHunter did not report any file containing the four specified types of evidentiary data, and we manually verified the analysis results for them. Our verification did not find any false negatives.

\myparatight{Privacy implications} We find that 5 apps save GPS locations on external storage; 8 apps save visited URLs on external storage; and 27 apps save text inputs on external storage. This result has serious privacy implications. Specifically, any app that has the READ\_EXTERNAL\_STORAGE and INTERNET permissions can read data from the external storage and send them to the Internet. If a user installs an app that saves sensitive data on the external storage, then other apps on the user's smartphone could monitor such data to compromise user privacy and security, even if the apps do not have permissions to access the sensitive data. For instance, an app can track user locations via monitoring GPS locations on the external storage; and an app can perform web tracking to profile a user via the visited URLs on the external storage. 

Via manual analysis, we found that one app, whose functionality is to backup contacts, saves a user's plaintext email address, phone number, and password to three separate files on the external storage.

\subsection{A Case Study}
\label{sec:evaluation_e_caseStudy}
We use a case study to demonstrate how a forensic investigator can use EviHunter to identify evidentiary data on an Android smartphone. Among the real-world apps that we analyzed, we found 133 apps could write GPS location and time data to a SQLite database with a static file path \emph{<internal storage>/databases/databases/ldata.db}. After manual analysis about these apps, we found that this database file is generated and accessed by a third-party advertisement library called \emph{airpush}. The package name for this library is \emph{com.yrkfgo.assxqx4}. The advertisement library registers a location listener.  Once a device's GPS location changes, the library will receive the GPS location, and then the library writes the GPS location as well as system time into the file \emph{<internal storage>/databases/databases/ldata.db}. The advertisement library uses GPS location to provide location-based advertisements. We speculate the reason that the library saves GPS locations in a local file system is to approximate a device's location when real-time GPS locations are unavailable.

\begin{figure}[]
\centering
\includegraphics[width=0.45\textwidth]{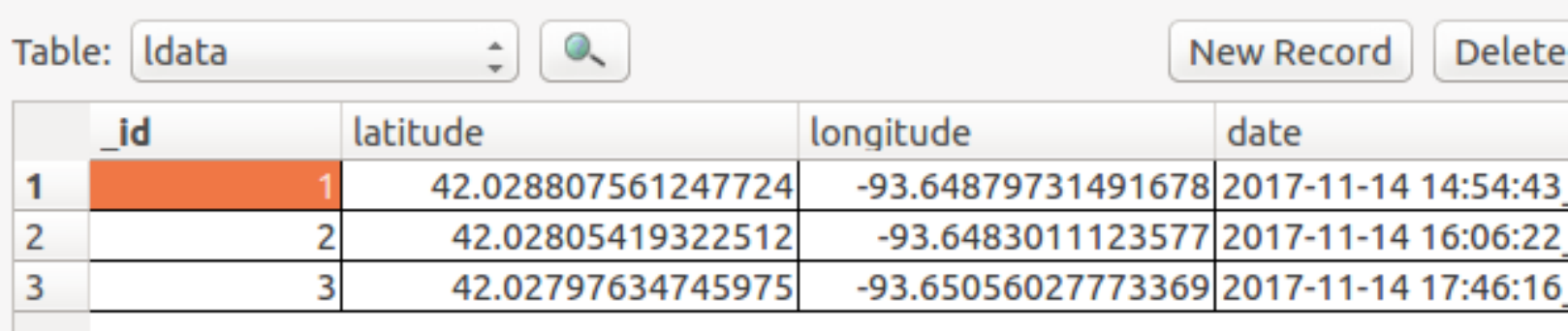}
\caption{GPS location and time data collected by the app in our case study.}
\label{fig:casestudy}
\end{figure}

We simulate a suspect's device using an Android smartphone and perform forensic investigation on the smartphone's file system. Specifically, we installed an app, whose package name is \emph{com.vijay.tamilrecipes}, on the smartphone. The app uses the \emph{airpush} library. One author of the paper walked around a building with the smartphone for a while, such that the app collected and stored GPS location and time data into the smartphone's file system. Then, we retrieved an image of the smartphone's file system. We matched the file system image against the AED we constructed in the previous section. In particular, we matched the file path \emph{/data/data/com.vijay.tamilrecipes/databases/databases/ldata.db} and our AED showed that this file path stores location and time data. Figure~\ref{fig:casestudy} shows a part of the database file. The columns of the database include \emph{\_id} (event id for location changes), \emph{latitude},  \emph{longitude}, and \emph{date}. A forensic investigator could use these GPS location and time data to assist crime investigation. For instance, a forensic investigator could use these data as a digital evidence to determine whether the suspect was at the crime scene or not when the crime happened. 

The total number of installations for these 133 apps is more than 10M. If a suspect has installed at least one of these apps on its Android smartphone, a forensic investigator could use the GPS location and time data saved by the app(s) as evidence for crime investigations.

\section{Discussions and Limitations}

EviHunter shares the limitations of many static data-flow analysis tools~\cite{Scandroid,kim2012scandal,AndroidLeaks,LeakMiner,CHEX,flowdroid,Amandroid,droidsafe,rDroid} for Android. 
Specifically, EviHunter over-approximates the system native methods and does not consider the developer-defined native methods. EviHunter only considers the reflective method calls that can be statically resolved. 
In general, EviHunter cannot handle dynamic class loading. For instance, an app could download a DEX file from Internet and execute it during runtime. Without runtime environment information, static program analysis cannot handle the loaded library. One possible way to mitigate this challenge is to download the dynamically loaded classes and statically analyze them together with the app.

EviHunter leverages IC3 and FlowDroid to support inter-component communications for the analysis of evidentiary data. However, EviHunter has basic support for inter-component communications with respect to file path propagation. In particular, if a dynamic file path includes intent in inter-component communication, EviHunter uses a regular expression <intent> as a part of the file path. 
Moreover, EviHunter has basic support for system APIs. In particular,  we manually summarize the data flows for system APIs that are related to file path constructions, string operations, and commonly used data structures. 
It would be an interesting future work to model frameworks. 
We note that the framework modeling technique proposed by DroidSafe~\cite{droidsafe} is insufficient for our file system forensics problem because of two reasons: 1) the technique is not scalable to real-world apps~\cite{starDroid}, and 2) the technique only captures the flows of sensitive data, but not file paths.

Another limitation revolves around matching file paths on a suspect's device against those in an AED. In particular, some dynamic file paths may not be successfully matched. For instance, suppose a file has a file name that completely consists of an intent. EviHunter will represent the file name as <intent>. However, since we do not produce any fine-grained patterns for the intent, it is hard to match the file on a suspect's device with the right dynamic file path in the AED. As a result, we will incorrectly treat the file as a file that does not include evidentiary data. We believe it is an interesting future work to extend our tool to support more fine-grained analysis for dynamic file paths. For instance, via inter-component communication analysis, we may uncover patterns within an intent. We can include these fine-grained patterns into the regular expression <intent> in a dynamic file path. When matching file paths, we can consider such patterns via regular expressions.

When our tool identifies files on a suspect's device that contain evidentiary data, a forensics investigator still needs to retrieve, decode, or reconstruct the evidentiary data from the files. If a file is a text file, SQLite database, or SharedPreferences, then retrieving data from the file is relatively easy. However, for a binary file or an encrypted file, a forensic investigator needs the format of the binary file or the encryption method to retrieve the evidentiary data. Our current tool does not support the format analysis of binary files. 
It is an interesting future work to extend EviHunter to track the format of a binary file.

\section{Related Work}
\label{sec:relatedwork}

\subsection{Digital Forensics for Android}
\subsubsection{Permanent-Storage Forensics}
Permanent-storage forensics for Android is still an underdeveloped research area.   
Most existing studies and tools on this topic simply leverage either manual analysis or keyword search. As a result, they can only analyze a small number of apps or construct an inaccurate AED.  
Our work represents the first one to perform automated permanent-storage forensic analysis for Android via program analysis. 

\myparatight{Manual analysis} Some studies~\cite{al2013skype,alyahya2017snapchat,anglano2014forensic,wu2017forensic} manually analyzed apps in order to construct an AED. Specifically, they install apps on an Android device or run the apps in a sandbox environment (e.g., Android Emulator~\cite{AndroidEmulator} and YouWave~\cite{YouWave}). Then, they retrieve an image of the file system from the device or the sandbox environment. Specifically, the file system image can be retrieved from a device using the Android Debug Bridge. The file system image can be either logical or physical, where a physical image could also include the deleted files that are not overwritten yet. By running apps under a sandbox environment,  researchers have control over the file system and main memory, so they can also retrieve images of the RAM and NAND flash memories.
After obtaining a file system image, they manually examine the files generated by the apps, e.g., analyzing the files under a directory \emph{/data/data/<package name>/files/}, where  \emph{<package name>} refers to the package name of the app to be analyzed. 

Since manual analysis is time-consuming, error-prone, and costly, these studies often only analyzed a very small number of apps. In particular, they often focused on instant messaging apps (e.g., WhatsApp~\cite{anglano2014forensic}, WeChat~\cite{wu2017forensic}) and maps navigation apps~\cite{maus2011forensic}. 
They found that instant messaging apps often save messages (i.e., a certain type of \emph{text input}) on the local file system, while maps navigation apps collect GPS location history and post on databases. %

\myparatight{Keyword search} 
Several commercial tools (e.g., Cellebrite UFED~\cite{UFED}, XRY~\cite{XRY}, and FTK~\cite{FTK}) are available to analyze the files on a device. Specifically, given a device, these tools first retrieve an image of the device's file system. Then, these tools provide Graphical User Interface (GUI) for forensic investigators to search files that could contain evidentiary data. However, the search is only performed by keyword matching or regular expression matching, which clearly has limitations. For instance, if a file contains GPS data but does not have the regular expressions or keywords such as GPS, latitude, or longitude, then the file will be incorrectly labeled as a file that does not contain evidentiary data. Indeed, studies~\cite{mfHans2017} showed that tools based on keyword matching can only identify a small fraction of files that could store evidentiary data.

\subsubsection{Memory Forensics} 
Several tools~\cite{DSCRETE,VCR,GUITAR,RetroScope}  were recently developed for memory forensics. For instance, DSCRETE~\cite{DSCRETE} automatically renders data structure contents in memory images. The intuition is that the app that defines a data structure often includes rendering and interpretation logic for the data structure, and DSCRETE leverages such logic. VCR~\cite{VCR} is a memory forensics tool that recovers all photographic evidence an Android device's camera produces. GUITAR~\cite{GUITAR} is a memory forensics tool to automatically reconstruct apps' GUIs from a memory image. However, GUITAR cannot  reconstruct an app's previous screens. RetroScope~\cite{RetroScope} addressed this limitation by leveraging two observations:  
(1) app data on previous screens stay longer than the corresponding GUI data structures in memory, and (2) each app can redraw its screens when receiving commands from the Android framework.  
These tools are different from ours as our tool is designed for permanent-storage forensics instead of memory forensics.

\subsubsection{Active Forensics}
Several studies~\cite{karpisek2015whatsapp, lee2009design, WebCapsule, xudynamic2018} proposed \emph{active forensics}.  Specifically, they proposed to design monitoring apps, which are installed on a device in advance. Then, the app can collect the forensic information when the device uses other apps. For instance,  DroidWatch~\cite{grover2013android} monitors the data collected and generated by other apps on the device through ContentObserver and Broadcast Receiver. Lee \textit{et al.}~\cite{lee2009design} designed a monitoring app that leverages system APIs to observe data including device ID, MAC address, and running processes. WebCapsule~\cite{WebCapsule} is a lightweight and portable forensic engine for web browsers. WebCapsule needs to be installed together with a web browser and records all non-deterministic inputs to the web rendering engine. The goal of WebCapsule is to reconstruct web security attacks, e.g., phishing attacks. 
A key challenge for active forensics is that it is almost impossible to install the monitoring app on a suspect's device before a crime happened. Our EviHunter is designed for \emph{passive forensics}, i.e., we analyze a suspect's device after a crime has happened. 

\subsection{Static Data Flow Analysis for Android}
Improving the security and privacy of Android has attracted lots of attention in the security community. 
A large number of tools have been developed to detect sensitive data flows between sources and sinks in Android apps. For instance, dynamic analysis tools include TaintDroid~\cite{taintdroid}, TaintART~\cite{taintart}, and Malton~\cite{malton}. Example static analysis tools include SCanDroid~\cite{Scandroid}, Scandal~\cite{kim2012scandal}, AndroidLeaks~\cite{AndroidLeaks}, LeakMiner~\cite{LeakMiner}, CHEX~\cite{CHEX}, FlowDroid~\cite{flowdroid}, AmanDroid~\cite{Amandroid}, DroidSafe~\cite{droidsafe}, R-Droid~\cite{rDroid}, IccTA~\cite{iccta}, and HornDroid~\cite{HornDroid}. Reaves \emph{et al.}~\cite{starDroid} performed a systematic comparative study about a large number of tools. Both dynamic analysis and static analysis can be used for permanent-storage forensics. For instance, Xu \textit{et al.}~\cite{xudynamic2018} designed a dynamic analysis tool by extending the Android platform's memory heap space to track information flow and report evidentiary data once a sink method is invoked. %

EviHunter leverages static analysis. Static analysis of Android apps relies on a comprehensive model to approximate Android's runtime behavior. 
CHEX~\cite{CHEX} was the first static analysis tool to consider different types of entry points of an Android app. 
FlowDroid~\cite{flowdroid} models an app's life cycle and creates a dummy main method for an app, which was also adopted by AmanDroid~\cite{Amandroid} and R-Droid~\cite{rDroid}. Many tools manually summarized the data flows for some commonly used system APIs  while ignoring the rest of them for efficiency consideration. Specifically, with the summarized data flows, a tool does not need to include the framework code, which is more efficient. DroidSafe~\cite{droidsafe} proposed a technique to model the Android framework, which was also used by R-Droid~\cite{rDroid}. However, Reaves \emph{et al.}~\cite{starDroid} found that DroidSafe is not scalable to real-world apps and fails to analyze them. Several tools~\cite{iccta,CHEX,flowdroid,Amandroid,droidsafe} also considered inter-component communications, which can increase the precision of the analysis.

A key difference between these static analysis tools and EviHunter is that EviHunter considers both the types of sensitive data and  file paths where the data are written to. Specifically, the tools that were designed to improve the security and privacy of Android focused on identifying what sensitive data (e.g., locations, contacts, device ID) are leaked to which types of sinks (e.g., network socket, file system, device's screen). Our EviHunter focuses on the file system as a type of sink and further identifies the files where data are written to. This key difference introduces several challenges: 1) we need to design a new tag that contains both evidentiary data and file path, 2) we need to design new propagation rules to update tags, especially for  file paths, and 3) we also need to identify new source methods for file paths. EviHunter addresses these challenges.

\section{Conclusion and Future Work}
\label{sec:conclusion}

In this work, we design EviHunter to automatically identify the files on an Android smartphone that could contain evidentiary data. EviHunter first builds an \emph{App Evidence Database} for a large number of apps via static data flow analysis. Then, EviHunter matches the files on a smartphone against those in the AED to identify the files containing evidentiary data. EviHunter extends the existing static analysis tools that were designed for security and privacy analysis of Android apps with respect to 1) tags associated with variables, 2) propagation rules to update tags, 3) data-flow summary of commonly used system APIs, and 4) sources and sinks. Our evaluations on both benchmark apps and real-world apps show that EviHunter can precisely and accurately identify the evidentiary data that an app could save to file systems and the files where the data are written to.  
 Future work includes modeling inter-component communications and frameworks for file path construction as well as extending EviHunter for fine-grained privacy analysis.

\begin{acks}
We would like to thank the anonymous reviewers for their insightful feedback. 
This work was funded by the Center for Statistics and Applications in Forensic Evidence (CSAFE) through Cooperative Agreement \#70NANB15H176 between NIST and Iowa State University, as well as partially by  NSF under grants No. CNS-1527579, CNS-1619201, and CNS-1730275, and Boeing Company.

\end{acks}

\balance
\bibliographystyle{ACM-Reference-Format}
\bibliography{RefPaper}

\end{document}